\documentclass[twocolumn]{aastex6}
\usepackage{graphicx}
\usepackage{epsfig}
\usepackage{amsmath, amssymb , amsthm, fouriernc}
\usepackage{threeparttable}

\newcommand{\source}{PSR~J1119--6127}
\newcommand{\src}{PSR~J1119--6127}

\newcommand{\nustar}{\textit{NuSTAR}}
\newcommand{\swift}{\textit{Swift}}

\newcommand{\xmm}{\textit{XMM-Newton}}
\usepackage{times}

\newcommand{\myemail}{rarchiba@physics.mcgill.ca}
\begin{document}

\title{A Magnetar-like Outburst from a High-B Radio Pulsar}

\author{R. F. Archibald,
 V. M. Kaspi,
 ~S.~P. Tendulkar,
 ~and ~P.~Scholz
}

\affil{Department of Physics and McGill Space Institute, McGill University, 3600 University St., Montreal QC, H3A 2T8, Canada\\Email: \myemail}

\begin{abstract}
Radio pulsars are believed to have their emission powered by the loss 
of rotational kinetic energy. By contrast,
magnetars show intense X-ray and $\gamma$-ray radiation whose luminosity
greatly exceeds that due to spin-down and is believed
to be powered by intense internal magnetic fields.
A basic prediction of this picture is that radio pulsars of
high magnetic field should show magnetar-like emission.
Here we report on a magnetar-like X-ray outburst from the radio 
pulsar PSR J1119$-$6127, heralded by two short bright X-ray bursts on 2016 July 27 and 28 \citep{klm+16,ykr16}. 
Using Target-of-Opportunity data from the {\it Swift} X-ray Telescope and
{\it NuSTAR}, we show that this pulsar's flux has brightened by a factor
of $>160$ in the 0.5--10~keV band, and its previously soft X-ray spectrum
has undergone a strong hardening, with strong pulsations appearing for the
first time above 2.5\,keV, with phase-averaged emission detectable up to 25 keV.
By comparing {\it Swift}-XRT and {\it NuSTAR} timing data with a pre-outburst
ephemeris derived from {\it Fermi} Large Area Telescope data, we find that the source
has contemporaneously undergone a large spin-up glitch of amplitude $\Delta\nu/\nu = 5.74(8) \times 10^{-6}$.  
The collection of phenomena observed thus far in this outburst strongly
mirrors those in most magnetar outbursts and provides an unambiguous
connection between the radio pulsar and magnetar populations.

\end{abstract}

\section{Introduction}
\label{sec:intro}

PSR J1119$-$6127 is a radio pulsar having spin period $P=0.407$~s,
discovered in the Parkes multibeam 1.4-GHz survey
\citep{ckl+00}.  The pulsar's $P$ and spin-down rate $\dot{P} = 4.0 \times 10^{-12}$
imply a characteristic age $\tau < 2$~kyr, making this object one of the youngest pulsars in
the Galaxy, consistent with an association with the supernova
remnant G292.2$-$0.5 \citep{cgk+01} at a distance of 8.4\,kpc \citep{cmc04}.
Those same spin parameters, under the assumption of
vacuum dipole spin-down, imply a dipolar surface
magnetic field $B = 4.1 \times 10^{13}$~G,
among the highest known among radio pulsars.  The
pulsar's spin-down luminosity is $\dot{E} = 2.3 \times 10^{36}$~erg~s$^{-1}$.

Past X-ray observations of the source \citep{gs03b,sk08,nkh+12}
have shown it to be a soft X-ray pulsar, with strong pulsations below
2.5~keV, and none seen above this energy.  This emission was well
described by a two-component model consisting of a power law of index $\sim$2.1, with a hot thermal
component of blackbody temperature $kT \simeq 0.2$~keV, high compared to
lower-field radio pulsars of comparable age
\citep[see also][]{km05,zkm+11,ozv+13}.  Radio pulse profile
changes, short radio bursts and unusual timing recoveries
were observed near epochs of glitches in this source
\citep{wje11,awe+15}, reminiscent of radio radiative behavior \citep[e.g.][]{crh+06} and
glitch recoveries \citep{dk14}
following magnetar outbursts.  The pulsar is also among the highest-B sources to have
been detected in $\gamma$-rays by {\it Fermi} \citep{pkh+11}.

The {\it Fermi} Gamma-ray Burst Monitor (GBM) and {\it Swift} Burst Alert Telescope
(BAT) both reported short magnetar-like bursts from PSR J1119$-$6127,
on 2016 July 27 \citep[UT 13:02:08;][]{ykr16} and 2016 July 28 \citep[UT 01:27:51;][]{klm+16},
respectively.  Immediately following the BAT burst,
the {\it Swift} X-ray Telescope (XRT) found a bright X-ray source
at the position of PSR J1119$-$6127 \citep{klm+16} with pulsations at the rotational period \citep{ave16}.  This suggests that this radio pulsar
has had a magnetar-like outburst, similar to the 2006 transition
of rotation-powered, but radio-quiet, pulsar PSR J1846$-$0258 \citep{ggg+08}. Interestingly, the radio pulsations from \source\ have disappeared \citep{bpk+16}.

We report here on our Target-of-Opportunity (ToO) X-ray observations of PSR J1119$-$6127 with
the {\it Swift}/XRT and {\it NuSTAR} during the first few days of the transition,
as well as on pre-outburst {\it Fermi}/LAT timing data.

\section{Observations \& Analysis}
\subsection{\swift-XRT \& \nustar\ Observations}
The {\it Swift}-XRT \citep{bhn+05} slewed to observe \source\ 62.8\,s after the BAT trigger \citep{klm+16}. XRT was operated in Photon Counting (PC) mode for this 2.2-ks observation (ObsID 00706396000, spanning 2016 July 28 01:28 -- 02:07 UT), and  in Windowed-Timing (WT) mode for the follow-up observations (ObsID 00034632001/2), spanning July 28 17:20 to July 29 03:11 UT and  July 31 04:20 to August 1 20:37 UT for exposures of 9.9 and 4.8\,ks, respectively. 

As the time resolution of PC mode is 2.5\,s (longer than the period of the pulsar), only WT mode observations, with a time resolution of $1.76\,$ms, were used in the timing analysis.

\nustar\ \citep{hcc+13} began ToO observations of \source\ at 2016 July 28, 23:05:12 UT yielding a total exposure time of 54.5\,ks  (ObsID 80102048002) partially overlapping with XRT observation 00034632001. The data from the two focal plane modules of \nustar\ are referred to hereafter as FPMA and FPMB. 

The \swift-XRT and \nustar\ data were processed with the standard \texttt{xrtpipeline} and \texttt{nupipeline} scripts, respectively, using \emph{HEASOFT} v6.17 and time corrected to the Solar System barycenter from the {\it Chandra} location  of \source{} \citep{gs03b}. 

For \swift, we selected only Grade 0 events for spectral fitting as other event grades are more likely to be caused by background events \citep{bhn+05}. \swift\ spectra were extracted from the selected regions using {\tt extractor}. Source photons were extracted from a 10-pixel radius circular region centered on \source\ with an annular background region with an inner and outer radius of 75 and 125 pixels, respectively. 

The WT observations had multiple soft X-ray bursts which appear in both the source and background regions that, from past experience, seem to be instrumental in origin. Hence, for the WT mode data, we excluded all photons below 0.7\,keV from our analysis.

For \nustar\ the source events were extracted within a 30-pixel (72\arcsec) radius around the centroid. Appropriate background regions were selected to be on the same detector as the source location. Spectra were extracted using the \texttt{nuproducts} script.

Using \texttt{grppha}, channels 0--70 ($<0.7$\,keV) and 700--1023 ($>7\,$keV) for \swift\ data and channels 0--35 ($<3$\,keV) and 1935--4095 ($>79\,$keV) for \nustar\ data were ignored and all good channels were binned to have a minimum of one count per energy bin. 

\subsection{{\it Fermi} Large Area Telescope Observations}
We downloaded Pass 8R2 events of {\it Fermi} Large Area Telescope (LAT) \citep{aaa+09a} all-sky survey observations from 2008 August 4 to 2016 July 30 from a one degree radius surrounding the {\it Chandra} position of \src\ and applied the recommended event selection. 
In the timing analysis, we used only photons having energy greater that 500~MeV based on the $\gamma$-ray pulse profile of \source\ \citep{pkh+11}. We corrected the arrival times of each photon to the Solar System barycenter using the {\tt tempo2} {\tt fermi} plug-in \citep{rkp+11}.

%Swift
\section{Results}

\subsection{Timing Analysis}
\label{sec:timing}
\begin{figure*}%[h!]
    \centering
    \includegraphics[width=\textwidth]{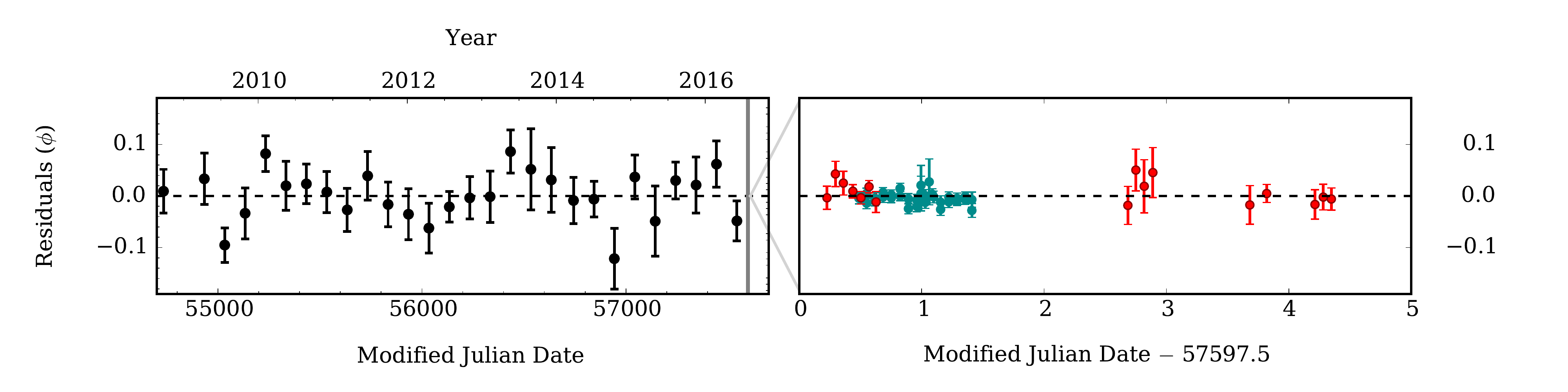}
    \caption{The left panel shows the LAT timing residuals of \src{} from MJD 54732 to 57544,  (2008 September 23 -- 2016 June 5), for the solution presented in Table~\ref{tab:timing}. The gray vertical bar shows the burst epoch. The right panel shows the residuals for the post-outburst timing solution. Black points are LAT, red are XRT, and cyan are {\it NuSTAR}.}
    \label{fig:resplot}
\end{figure*}

Times-of-arrivals (TOAs) of $\gamma$-ray pulses were extracted using a maximum likelihood (ML) method, described in \cite{lrc+09}. We extracted a TOA from photons collected in every 100 day span as a trade-off between TOA spacing and precision. We extended the  ephemeris presented by \citet{awe+15} using the LAT detected photons until the GBM-detected burst \citep{ykr16}. We present a phase-coherent ephemeris in Table~\ref{tab:timing} and the timing residuals in the left panel of Figure~\ref{fig:resplot}.

To determine an ephemeris for the post-outburst {\it Swift} and {\it NuSTAR} observations, we folded the soft X-ray photons ($<10$\,keV) from each observation, starting with the ephemeris from the LAT observations and extracted TOAs from each orbit using the ML method. As there is no apparent evolution in the pulse profiles over the $<10\,$keV energy band, the offset between TOAs from both telescopes should be minimal. We then used the {\tt tempo2} timing software package \citep{hem06} to fit the TOAs.

It is apparent that the LAT ephemeris did not accurately describe the post-outburst TOAs and requires a change in the spin frequency. Due to the long integration times required to extract a TOA from LAT, we are unable to constrain the exact glitch epoch; for this analysis, we have fixed the glitch epoch to the time of the first GBM-detected burst \citep{ykr16} and fitted for a glitch in spin frequency and frequency derivative.

We measure a spin-up glitch with $\Delta\nu=1.40(2)\times10^{-5}$\,Hz  and $\Delta\dot{\nu}=-1.9(5)\times10^{-12}\,$Hz s$^{-1}$. We caution that this represents a snapshot of the  frequency evolution, and that following glitches in magnetars, complex recoveries are often observed \citep[e.g.][]{dkg08, dk14}.

\begin{table}
    \begin{center}
    \caption{Phase-Coherent Ephemeris for \src{}.}
    \label{tab:timing}
    \begin{tabular}{ll}
    \hline
    \hline
    \multicolumn{2}{c}{{\it Fermi} LAT Ephemeris} \\
    Dates (MJD)         & 54732.82--57544.08\\
    Dates               & 2008 Sept 23 -- 2016 June 5 \\
    Epoch (MJD)         & 56264.00000\\
    $\nu\;$ (s$^{-1}$)            &  2.442 579 294 0(9)\\
    $\dot{\nu}\;$ (s$^{-2}$)            &$-$2.390 210(4)$\times 10^{-11}$\\
    $\ddot{\nu}\;$ (s$^{-3}$)            & $5.68(3)\times 10^{-22}$ \\
    $\dddot{\nu}\;$(s$^{-4}$)      & $ -1.46(9)\times 10^{-30}$  \\
    $\nu^{(4)}$$\;$(s$^{-5}$)      & $ -3.1(7)\times 10^{-38}$  \\
    $\nu^{(5)}$$\;$(s$^{-6}$)      & $ 1.8(1)\times 10^{-45}$  \\
    $\nu^{(6)}$$\;$(s$^{-7}$)      & $ 3(1)\times 10^{-53}$  \\
    RMS residual (ms) & 15.5\\
    RMS residual (phase) & 0.037\\
    $\chi^2_\nu$/dof & 1.08/20 \\
    \hline
    \multicolumn{2}{c}{Post-Outburst Ephemeris} \\
    Dates (MJD)         & 57597.72--57601.85\\
    Dates               & 28 July --1 Aug 2016 \\
    Epoch (MJD)         & 57600.\\
    $\nu\;$ (s$^{-1}$)            &  2.439 837 34(8)\\
    $\dot{\nu}\;$ (s$^{-2}$)            &$-$2.57(5)$\times 10^{-11}$\\
    RMS residual (ms) & 4.22\\
    RMS residual (phase) & 0.001\\
    $\chi^2_\nu$/dof & 0.74/46 \\
    \hline
    \multicolumn{2}{c}{Glitch Parameters} \\
    Glitch Epoch (MJD), fixed         & 57596.547\\
    $\Delta\nu\;$ (s$^{-1}$)            &  $1.40(2)\times 10^{-5}$\\
    $\Delta\dot{\nu}\;$ (s$^{-2}$)           &  $-1.9(5)\times 10^{-12}$\\
    \hline
    \hline
    \newline
    \end{tabular}
    \newline
    Note: Figures in parentheses are the  1$\sigma$ \textsc{tempo2} uncertainties in the least-significant digits quoted. The source location was fixed at the {\it Chandra} position.
    \end{center}
\end{table} 

\begin{figure}
  \center
  \includegraphics[width=0.48\textwidth]{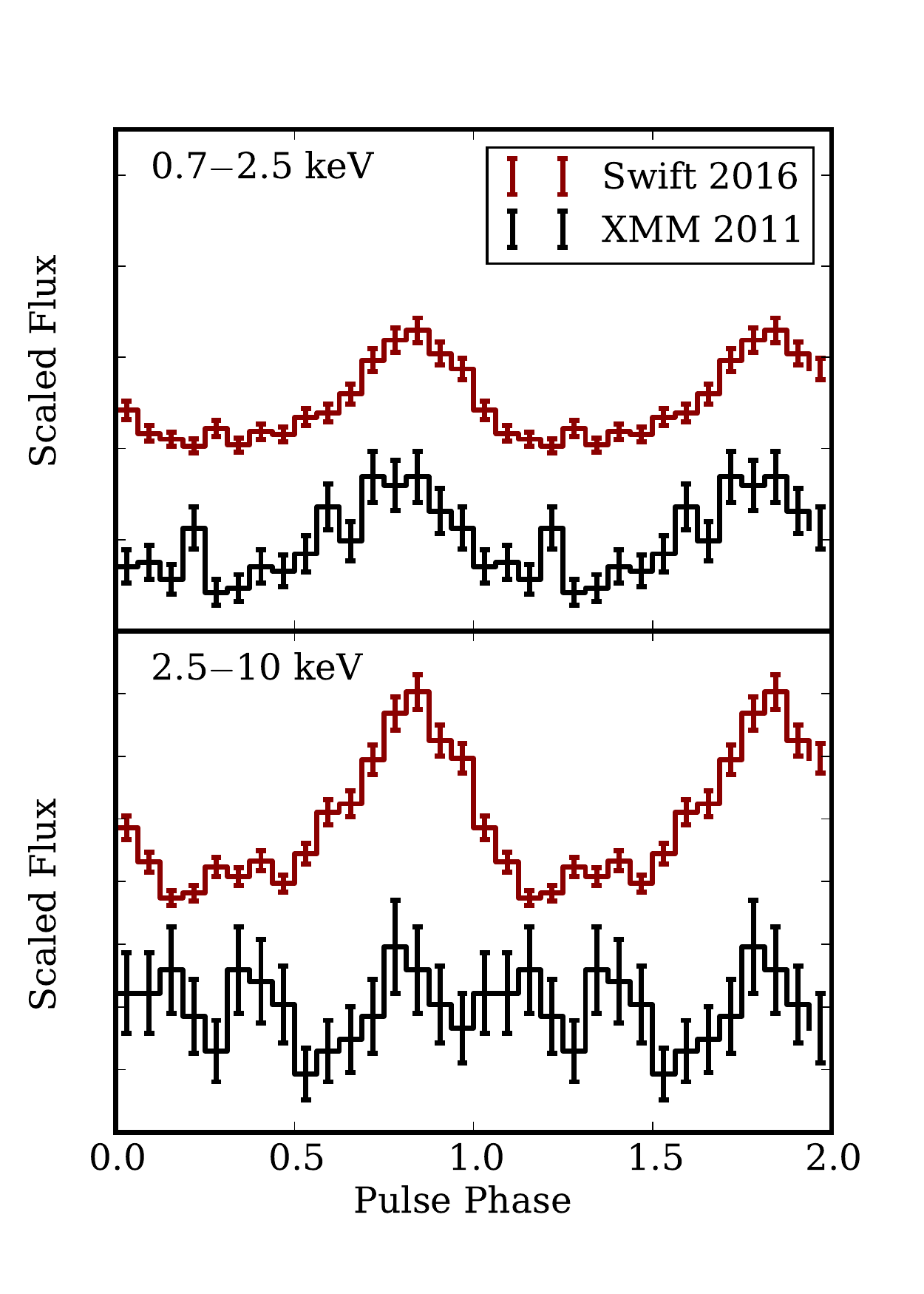}
  \caption{Pre- and post-outburst X-ray pulse profiles of \source\ from \swift-XRT (00034632001, red) and 2011 \xmm\ data \citep{nkh+12} (black) in the soft (0.7--2.5\,keV, top panel) and hard (2.5--10\,keV, bottom panel). The profiles have been arbitrarily offset vertically for clarity and aligned in phase based on the soft X-ray profile.}
  \label{fig:profs}
\end{figure}

In Figure~\ref{fig:profs} we show the soft (0.7--2.5\,keV), and hard (2.5--10\,keV) X-ray pulse profiles from the radio pulsar state in 2011 \citep{nkh+12}, and the magnetar-like state in 2016. In the soft band, the root-mean-squared (RMS) pulse fraction has increased from 38(3)\% to 71(4)\%, while the pulse shape has remained similar. In the hard band, the RMS pulse fraction went from $<10\%$ to 56(3)\% pulsed.

\subsection{X-ray Spectroscopy}
\label{sec:spectroscopy}
All X-ray spectra were fit using \texttt{XSPEC} v12.9.0 \citep{arn96} with a common value for hydrogen column density ($N_\mathrm{H}$). Magnetar spectra are typically described with an absorbed blackbody plus power-law model, which we use here.  However, the independent \swift-XRT observations 00706396000 and 00034632002 were fit with a simple absorbed blackbody model between 0.7--7\,keV as there is no constraint on the power law without the \nustar\ spectrum and there is little power-law contribution below 7\,keV. 

We used Cash statistics \citep{cas79a} for fitting and parameter estimation of the unbinned data. $N_\mathrm{H}$ was fit using \texttt{wilm} abundances and \texttt{vern} photoelectric cross-sections. The normalizations of \nustar\ FPMB and \swift-XRT spectra were allowed to vary with respect to that of the \nustar\ FPMA spectrum.

The hard X-ray tail seen above 8\,keV in the PC mode observation (00706396000) may be caused by contamination due to short temporally unresolved X-ray bursts. Magnetar-like bursts are intrinsically harder than the average spectrum and the high count-rate leads to pile-up effects within the 2.5-s CCD readout time \citep[e.g.][]{sk11}. These pile-up effects are not mitigated by standard techniques such as the removal of the central bright region, as they are bunched temporally rather than spatially. If the hard X-ray tail were indeed real, it would have needed to fade by $\sim2$ orders of magnitude in the day before the {\it NuSTAR} pointing to be consistent with the measured hard X-ray flux, while the blackbody temperatures measured on the two epochs are consistent with slow cooling. Hence, to fit the average spectra, we truncate the 00706396000 spectra above 7\,keV.

\begin{figure}
  \center
  \includegraphics[width=0.48\textwidth]{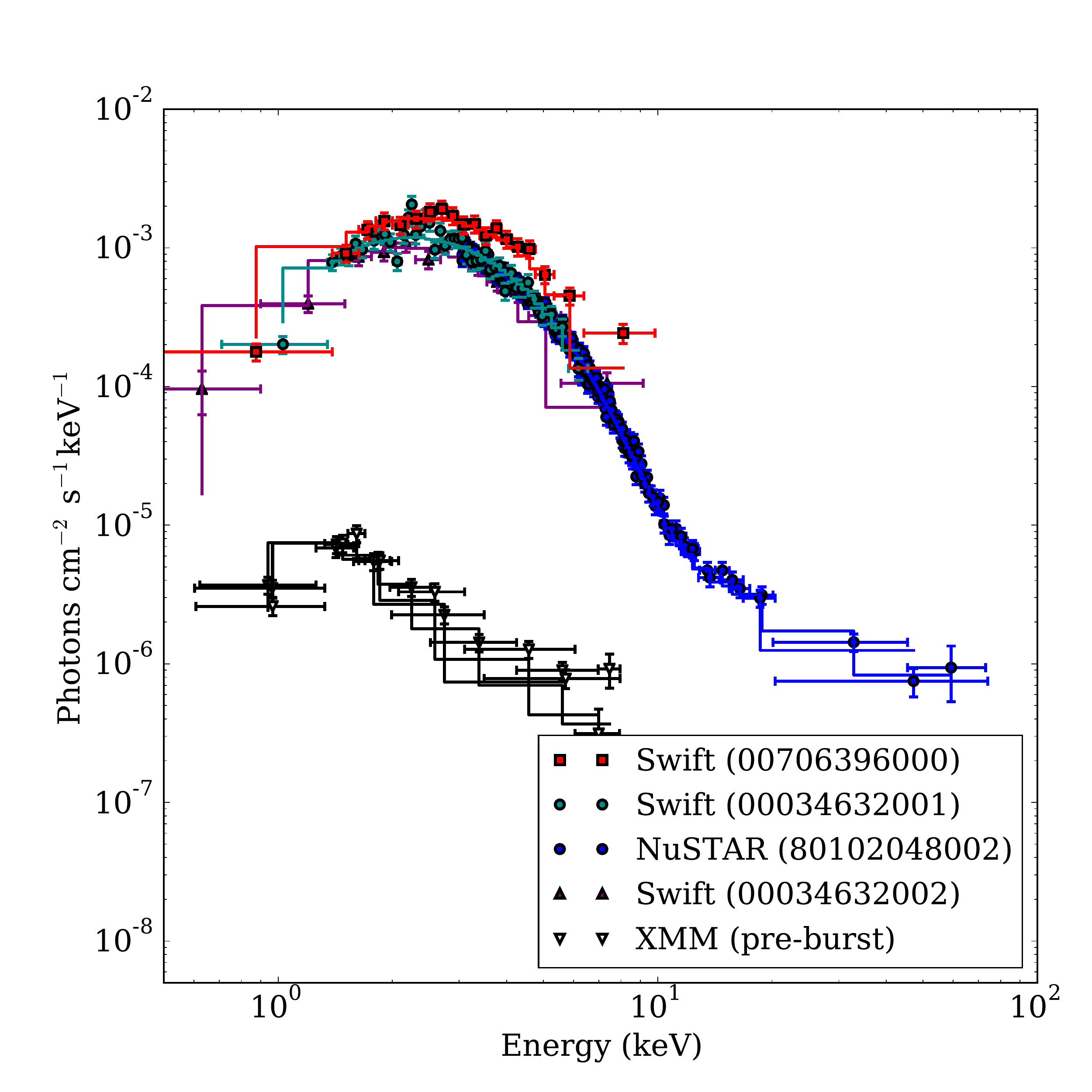}
  \caption{X-ray spectra of \source. The data are as follows: pre-outburst \xmm\ spectrum (inverted black triangles), \swift-XRT spectrum at burst (red squares), \swift-XRT and \nustar\ spectrum one day after the burst (green and blue circles, respectively), and a \swift-XRT spectrum three days after the burst (purple triangles). The flux increase above 8\,keV in the \swift-XRT burst spectrum is likely due to pileup  (see Section~\ref{sec:spectroscopy} for details). For clarity, the \nustar\ spectra from FPMA and FPMB are combined.}
  \label{fig:1119_all_spec}
\end{figure}

Figure~\ref{fig:1119_all_spec} shows the current \swift+\nustar\ spectral fits in comparison with a pre-burst \xmm\ spectrum \citep[from ][]{nkh+12}. Table~\ref{tab:nustar_spectra} details the parameter values with 90\% confidence error bars.

Immediately after the burst, we measure a blackbody temperature $kT=1.10(6)$\,keV, slightly decreasing to $0.96(1)$\,keV and $0.93(1)$\,keV in the follow-up spectra. This is substantially higher than the pre-burst blackbody temperature of 0.21(4)\,keV \citep{nkh+12}. In the \nustar\ spectra, we also measure a hard power law with photon index $\Gamma=1.2(2)$ that is marginally harder than the pre-burst  value $\Gamma_\mathrm{PSR}=2.1(8)$ \citep{nkh+12}. We also note that \citet{nkh+12} and \citet{sk08} measured the power law emission from the pulsar wind nebula (PWN) around \source\ to have $\Gamma_\mathrm{PWN} = 1.1-1.4$, close to the current hard power law index, but with a flux of $\sim2 \times 10^{-14}\,\mathrm{erg\,cm^{-2}\,s^{-1}}$, almost three orders of magnitude fainter.

\begin{deluxetable}{llc}
%\centering
\tablecolumns{3}
\tablecaption{Spectral Fits to NuSTAR and Swift-XRT Data\label{tab:nustar_spectra}}
\tablewidth{0pt}
\tabletypesize{\footnotesize}
\tablehead{\colhead{Component}& \colhead{Parameter} & \colhead{Value}}
\startdata
\texttt{tbabs}    & $N_\mathrm{H}$ ($10^{22}\,\mathrm{cm^{-2}}$) & $1.2\pm0.1$\\
\sidehead{\swift-XRT 00706396000 (\texttt{tbabs*bbody})}
\texttt{bbody}    & $kT_\mathrm{BB}$ (keV) & $1.10\pm0.06$ \\
                   $\mathrm{C-Stat}/\mathrm{dof}$ & & 333.82/413 \\
                   goodness\tablenotemark{a} & & $12\%$ \\
                   Flux (0.5--10\,keV)\tablenotemark{b} & & $4.1\pm0.1$ \\
                   $L_X$ (0.5--10\,keV)\tablenotemark{c} & & 3.5 \\
\sidehead{\vspace{-0.4cm} \nustar\ 80102048002 + \swift-XRT 00034632001} 
\sidehead{\texttt{const*tbabs*(bbody+powerlaw)}} 
\texttt{const}    & $C_{\mathrm{FPMB}}$\tablenotemark{d} & $1.01\pm0.02$  \\
                  & $C_\mathrm{XRT}$\tablenotemark{d}   & $0.94\pm0.04$ \\
\texttt{bbody}    & $kT_\mathrm{BB}$ (keV) & $0.96\pm0.01$ \\
\texttt{powerlaw} & $\Gamma$ & $1.2\pm0.2$  \\
                   $\mathrm{C-Stat}/\mathrm{dof}$ & & 2133.1/2327 \\
                   goodness\tablenotemark{a} & & $20\%$ \\
  Flux (0.5--10\,keV)\tablenotemark{b} & & $2.7\pm0.1$ \\
                   $L_X$ (0.5--10\,keV)\tablenotemark{c} & & 2.3 \\ 
  Flux (3--79\,keV)\tablenotemark{b} & & $1.9\pm0.1$  \\ 
                   $L_X$ (3--79\,keV)\tablenotemark{c} & & 1.6 \\
\sidehead{\swift-XRT 00034632002 (\texttt{tbabs*bbody})}
\texttt{bbody}    & $kT_\mathrm{BB}$ (keV) & $0.93\pm0.06$ \\
                   $\mathrm{C-Stat}/\mathrm{dof}$ & & 304.7/355 \\
                   goodness\tablenotemark{a} & & $34\%$ \\
                   Flux (0.5--10\,keV)\tablenotemark{b} & & $2.1\pm0.2$ \\
                   $L_X$ (0.5--10\,keV)\tablenotemark{c} & & 1.8 \\
\enddata
\tablenotetext{a}{Percentage of C-Stat statistic simulation trials from model parameters that are less than the fit statistic.}
\tablenotetext{b}{Unabsorbed flux in units of $10^{-11}\,\mathrm{erg\,cm^{-2}\,s^{-1}}$.}
\tablenotetext{c}{X-ray luminosity assuming isotropic emission at a distance of 8.4\,kpc in units of $10^{35}\,\mathrm{erg\,s^{-1}}$.}
\tablenotetext{d}{Cross-normalization constants w.r.t. \nustar\ FPMA.}
\end{deluxetable}

\section{Discussion}
\label{sec:discussion}

The outburst from PSR J1119$-$6127 is observationally very similar to those
seen in magnetars.
The phenomenology of magnetar outbursts is rich \citep[see][for a review]{re11}, but with
established commonalities, practically
all of which are observed in the PSR J1119$-$6127 event.
A hallmark of magnetar outbursts are short-duration ($<1$-s) 
hard X-ray bursts, as reported for PSR J1119$-$6127 \citep{klm+16,ykr16}. 
The large flux enhancement, here by a factor of $>160$, is commonly seen in magnetar outbursts,
notably those in which the quiescent luminosity is below 
$\sim 10^{33}$~erg~s$^{-1}$ \citep[e.g.][]{sk11,kkp+12}.  
Moreover, the spectral hardening we report is classic for magnetar outbursts \citep{re11}, as
are timing anomalies, most often spin-up glitches \citep{dkg08,dk14}.  
Overall, the PSR J1119$-$6127 event is clearly magnetar-like.

Most similar to the
PSR J1119$-$6127 event is the 2006 magnetar-like outburst of PSR J1846$-$0258
\citep{ggg+08}.  The latter is also a young ($\tau < 1000$~yr), high-B  
($B=5 \times 10^{13}$~G) rotation-powered pulsar, albeit radio undetected \citep{aklm08}.  Next we compare
this object's 2006 outburst with the event studied here.

One difference between the outbursts is their energetics.  
For PSR J1119$-$6127, the 0.5--10-keV flux as measured in the joint
{\it Swift} and {\it NuSTAR} observation was $4.1(1) \times 10^{-11}$~erg~s$^{-1}$~cm$^{-2}$.
For a distance of 8.4~kpc, and assuming
isotropic emission, this implies a luminosity of $3.5 \times 10^{35}$~erg~s$^{-1}$,
or 0.1$\dot{E}$.  This represents an increase over the quiescent
value in this band of a factor of over $\sim160$.
Including the flux from 10~keV extrapolated to the top of the {\it NuSTAR} band
increases this value by $\sim$20\%.  The
efficiency for conversion of $\dot{E}$ to {\it Fermi}-band $\gamma$-ray emission,
at least in quiescence, was estimated by
\citet{pkh+11} to be 0.23.  With a comparable amount
of energy suddenly appearing in X-rays, the {\it Fermi}-band emission may have been affected during
this outburst. However, the normally low {\it Fermi}/LAT count rate 
(see \S\ref{sec:timing}) requires multiple weeks of integration for a detection, hence
a short-term anomaly may be undetectable.

By contrast, for PSR J1846$-$0258, 
in {\it Chandra} observations made within one week of its first detected
magnetar-like burst, the unabsorbed flux in the 0.5--10-keV
band was $4.0^{+1.6}_{-0.9} \times 10^{-11}$~erg~s$^{-1}$~cm$^{-2}$
\citep{ggg+08}.  For isotropic emission and a distance of 6~kpc
\citep{lt08}, this implies a luminosity of $1.7 \times 10^{35}$~erg~s$^{-1}$,
or 0.02$\dot{E}$.  This represented an increase of a factor of $7.7^{+3.9}_{-1.8}$ over
the quiescent value.  As this outburst had no realtime trigger, 
it is possible that initially the pulsar
brightened more.  However the pulsed flux measured by {\it RXTE}
at the initial burst epoch was comparable a week later. 
Hence the increase in flux was probably not much larger than a factor of $\sim$10, never
exceeding a few percent of $\dot{E}$. 
Thus energetically, 
the PSR J1119$-$6127 outburst is far larger than that in PSR J1846$-$0258. 

 The spectral evolution in the two events also differed.
In quiescence, PSR J1846$-$0258's X-ray spectrum is well described by a simple
power law of index 1.1(1) \citep{ggg+08}, very different
from the soft quiescent spectrum PSR J1119$-$6127.  The latter, well described
by a blackbody of $kT = 0.21(4)$~keV and power law of index 2.1(8)
\citep{sk08,nkh+12}, made this pulsar the youngest with
detected thermal emission.  It was also one of the hottest and most luminous thermal
emitters even among rotation-powered pulsars of similar age \citep{ozv+13}.
This emission was also noted to be unusual for its high pulsed fraction
\citep{nkh+12}.

During outburst, PSR~J1846$-$0258 developed a bright soft component, whereas the harder power-law
spectrum remained unchanged, apart from an increase in normalization by $\sim$35\%
above 10~keV \citep{kh09}.  The soft component was described
as having a power-law spectrum of index 1.9(1) \citep{ks08b,ggg+08}. 
By contrast, the spectrum of PSR J1119$-$6127 has undergone
a radical hardening, with $kT$ increasing from 0.21(4)\,keV to 1.10(6)\,keV.
Nevertheless, the net effect in both outbursts was a transition to
a spectrum very similar to those of bright magnetars.

It is also interesting to compare the outburst timing anomalies.
In PSR J1846$-$0258, it suffered a
sudden spin-up having $\Delta\nu / \nu \simeq 3 \times 10^{-6}$, 
followed by a large increase in $\dot{\nu}$ yielding a strong over-recovery of the glitch 
\citep{lkg10,kh09,lnk+11}.
The net long-term effect was a spin-{\it down}, accompanied by a change
in braking index and a long-term enhancement in timing noise. 
While it is too early to know the post-outburst timing evolution in PSR~J1119$-$6127, from
our timing analysis, we find that the pulsar had a similar-sized
spin-up glitch with $\Delta\nu / \nu \simeq 5.8 \times 10^{-6}$. 
Presently any increase in spin-down rate is modest compared to some glitch recoveries
in magnetars and certainly compared to that following the 2006 PSR J1846$-$0258 glitch. However, greater
evolution may yet be detected.

In young radio pulsars like PSR J1119$-$6127, 
hard X-ray emission is thought to arise
in the context of outer gap models \citep[e.g.][]{wtk13}
from synchrotron radiation from secondary electron/positron
pairs produced by inward propagating curvature radiation
$\gamma$-rays. 
As discussed by \citet{pkh+11}, in PSR J1119$-$6127, the X-ray/$\gamma$-ray phase offset,
together with the single-peak
morphology of the $\gamma$-ray pulse, are well explained in
outer gap models.  
The luminosity of both the X-ray and $\gamma$-ray emission 
in this picture must be bounded by the spin-down power.

The increase in X-ray luminosity particularly in the hard
X-ray range during the outburst of PSR J1846$-$0258
was argued by \citet{kh09} to be plausibly due to the above-described
rotation-powered outer-gap emission, enhanced by particle
injection due to perhaps to crust cracking that occurred
at the glitch, 
reasonable given the lack of evidence for the hard X-ray luminosity
exceeding more than a few percent of $\dot{E}$.

The new hard X-ray emission component in PSR J1119$-$6127
could have an outer-gap origin as well, but
the large luminosity
rise to within 0.1$\dot{E}$ in the X-ray band alone may be difficult to accommodate in such
a picture, 
and might require a commensurate increase in $\gamma$-ray luminosity, impossible
given the available $\dot{E}$ energy budget.

Alternatively, the hard X-ray emission may be magnetar-like.
The origin of the bright hard X-ray component in magnetars has been argued to be a decelerating electron/positron
flow in the closed magnetosphere, in the higher altitude regions of 
large magnetic loops \citep{bel13}.  This emission is
powered ultimately by the internal stellar magnetic field and is not
limited by $\dot{E}$.
In this interpretation,
the pulsar suffered an instability such that a significant
twist in its field lines occurred, with highly relativistic particles
($\gamma >> 10$) injected near the star where $B >> B_{QED} = 4.4 \times 10^{13}$~G.
If this is origin of the hard X-rays in PSR J1119$-$6127,
then the true internal field of this pulsar
is far higher than is inferred from its dipole component. This would support
the argument for additional non-dipolar field components in apparently
low-magnetic-field magnetars \citep{ret+10,skc14}. One way to test
this explanation is through
modelling of the phase-resolved hard X-ray spectrum.
This can yield constraints on the geometry of the emission region
\citep[e.g.][]{hbd14}.  Such constraints could then be compared
with similar ones from radio polarimetry \citep{wje11}
and/or modelling of the $\gamma$-ray light curve \citep{pkh+11}.

Importantly, PSRs~J1846$-$0258 and now J1119$-$6127 are the only two rotation-powered pulsars
to have exhibited radiative changes at glitch epochs; this must be a consequence
of their high spin-inferred $B$.  
Indeed no X-ray enhancement was seen in {\it Chandra}
observations made
3.5 days following a large ($\Delta\nu/\nu = 3 \times 10^{-6}$) spin-up glitch in the lower-field
($B=3.4 \times 10^{12}$~G)
Vela radio pulsar \citep{hgh01}. 
On the other hand, there was no evidence for an X-ray enhancement in PSR J1846$-$0258 near the epoch of 
a much smaller glitch having $\Delta\nu/\nu = 2.5 \times 10^{-9}$ \citep{lkgk06}, 
nor in previous glitches ($\Delta\nu/\nu = 2.9 \times 10^{-7}$ and $4.1 \times 10^{-6}$ in 2004 and 2007, respectively) 
in PSR J1119$-$6127 \citep{wje11,awe+15}, although prompt X-ray observations were not performed in those cases.
Moreover, multiple sizeable glitches in {\it bona fide} magnetars have been unaccompanied by radiative changes
\citep{sak+14,dk14}.  This may indicate that an independent parameter 
such as the crustal depth of the glitch location plays a role in the radiative detectability
of high-B neutron-stars at glitch epochs \citep[see, e.g., ][]{ec89a,let02}.

The possibility of a magnetar-like outburst from a high-B
radio pulsar was discussed by \citet{km05}, who also suggested the possibility of 
radio emission from magnetars prior to its discovery by \citet{crh+06}.  
\citet{pp11a} and \citet{pp11b} provided theoretical groundwork for the
hypothesis, and 
magnetothermal modelling such as that by \citet{vrp+13} have further developed these ideas,
which are now on solid observational ground.  
Other high-B radio pulsars like PSRs J1718$-$3718 \citep{zkm+11} and J1734$-$3333 \citep{ozv+13} 
seem likely to also undergo a magnetar-like transition in coming years.

\smallskip
\noindent {\it Acknowledgements:}
The authors thank the operations teams of {\it NuSTAR}, particularly Karl Forster, and {\it Swift} for their speed and flexibility scheduling these observations.  We thank {\it Fermi}-LAT Collaboration for the public data and tools used in this work.  This work made use of data from the {\it NuSTAR} mission, a project led by the California Institute of Technology, managed by the Jet Propulsion Laboratory, and funded by the NASA. We acknowledge the use of public data from the {\it Swift} data archive.
R.F.A. acknowledges support from an  NSERC CGSD. V.M.K. receives support from an NSERC Discovery Grant, an Accelerator Supplement and from the Gerhard Herzberg Award, an R. Howard Webster Foundation Fellowship from the Canadian Institute for Advanced Study, the Canada Research Chairs Program, and the Lorne Trottier Chair in Astrophysics and Cosmology. S.P.T acknowledges support from a McGill Astrophysics postdoctoral fellowship. P.S. acknowledges support from a Schulich Graduate Fellowship.

\end{document}